\documentclass[pra, twocolumn, superscriptaddress, groupaddress, floatfix]{revtex4-2}
\usepackage{slashed}
\usepackage{graphicx}
\usepackage[colorlinks=true,linkcolor=blue,citecolor=blue,urlcolor=blue]{hyperref}
\usepackage{amsmath}
\usepackage{amsfonts}
\usepackage[toc]{appendix}
\usepackage{txfonts}
\usepackage[british]{babel}
\usepackage{times}
\usepackage[none]{hyphenat}

\everymath{\displaystyle} 
\newcommand{\pr}[1]{\ensuremath{\left[#1\right]}} 
\newcommand{\pc}[1]{\ensuremath{\left(#1\right)}} 
\newcommand{\px}[1]{\ensuremath{\left\lbrace#1\right\rbrace}} 
\newcommand{\bra}[1]{\ensuremath{\left\langle#1\right\vert}} 
\newcommand{\ket}[1]{\ensuremath{\left\vert#1\right\rangle}} 
 
\newcommand{\av}[1]{\ensuremath{\left\langle#1\right\rangle}} 

\newcommand{\eff}[0]{\text{eff}}
\newcommand{\im}[0]{\text{i}}

\usepackage{array}
\newcolumntype{M}[1]{>{\centering\arraybackslash}m{#1}}

\begin{document}

\title{Floquet engineering of continuous-time quantum walks: towards the simulation of complex and next-to-nearest neighbor couplings}
\author{Leonardo Novo}
\affiliation{Centre for Quantum Information and Communication, \'Ecole Polytechnique de Bruxelles, Universit\'e Libre de Bruxelles, Belgium}
\author{Sofia Ribeiro}
\affiliation{Joint Quantum Centre (JQC) Durham-Newcastle, Department of Physics, Durham University, United Kingdom}

\begin{abstract}
The formalism of continuous-time quantum walks on graphs has been widely used in the study of quantum transport of energy and information, as well as in the development of quantum algorithms. In experimental settings, however, there is limited control over the coupling coefficients between the different nodes of the graph (which are usually considered to be real-valued), thereby restricting the types of quantum walks that can be implemented. In this work, we apply the idea of Floquet engineering in the context of continuous-time quantum walks, i.e., we define periodically-driven Hamiltonians which can be used to simulate the dynamics of certain target continuous-time quantum walks. We focus on two main applications: i) simulating quantum walks that break time-reversal symmetry due to complex coupling coefficients; ii) increasing the connectivity of the graph by simulating the presence of next-to-nearest neighbor couplings. Our work provides explicit simulation protocols that may be used for directing quantum transport, engineering the dispersion relation of one-dimensional quantum walks or investigating quantum dynamics in highly connected structures. 
\end{abstract}
\maketitle
\section{Introduction}
Over the past decades, many efforts have been made in the study of quantum walks \cite{kempe2003QWreview}. Apart from describing fundamental physical processes such as the quantum transport of mass and energy, quantum walks also provide a framework for developing quantum state transfer protocols \cite{bose2003quantum,christandl2004perfect, kendon2011perfect} or quantum algorithms \cite{ambainis2003quantum}. This has motivated several experimental realizations of quantum walks, both in the discrete-time \cite{PRA48.1687(1993), ProcACM50(2001)} or continuous-time setting \cite{PRA58.915(1998)}, in a wide range of experimental platforms such as optical lattices \cite{PRA66.052319(2002), Science325.174(2009)}, ion traps\cite{PRL103.183602(2009), PRL104.100503(2010)}, NMR \cite{PRA72.062317(2005),PRA67.042316(2003)} or photonic devices \cite{ flamini2018review, szameit_review}. 

In this work, we focus on continuous-time quantum walks (CTQWs), where a quantum particle or excitation propagates in a network of coupled nodes via continuous-time evolution. Such a general model finds widespread applications in condensed matter physics as well as quantum information and computation. For one-dimensional chains, this model can be used to study, for example, the propagation of a single spin excitation in a spin chain \cite{bose2003quantum}, or localization phenomena in disordered media  \cite{PR109.1492(1958)}. More complicated highly connected networks have been considered in a variety of contexts, such as in models of energy transport in specific biological systems \cite{qbio1,qbio2,qbio3}, in the analysis of quantum walks on complex networks \cite{muelken2011, biamonte2019complex} or in spatial search algorithms \cite{childs2004spatial,chakraborty2016spatial,optimalityspatialsearch}.  

In most of the applications of CTQWs, the Hamiltonian is assumed to have real-valued parameters. This implies that time-reversal symmetry is respected, i.e., the probability for a particle to be transferred from one node to another at time $t=T$ is the same as at $t=-T$. However, complex-valued coupling parameters, which appear naturally in the description of charged particles in a magnetic field \cite{hofstadter}, can break time-reversal symmetry, offering the possibility of biasing the direction of propagation of the quantum walk as well as enhancing or suppressing quantum transport between different nodes \cite{ScienRep3.2361(2013), PRA93.042302(2016)}.
In an experimental setting, however, tuning coupling coefficients to arbitrary complex values are, in general, not possible, especially in systems of neutral particles such as photons or ultracold atoms. Furthermore, in physical systems the coupling strength between different nodes typically decays with the distance, thereby restricting the quantum walks on graphs that can be implemented experimentally or observed in nature. One possibility to partially overcome the latter restriction was proposed in Ref.~\cite{ehrhardt2020exploring} where by coupling internal degrees of freedom of the particles (in this case, the photon polarization) with spatial degrees of freedom it was possible to implement quantum walks in certain highly connected graphs.  

In our work, we exploit the idea of Floquet engineering \cite{AdvPhys64.139(2015)} in the context of CTQWs and demonstrate that this versatile technique can be used to create complex coupling coefficients in quantum walks as well as increase the connectivity of the graph where the quantum walk takes place. The idea behind Floquet engineering is to design a time-periodic Hamiltonian in a specific way, such that the dynamics is approximately described by an effective Hamiltonian with the characteristics that we would like to simulate. This effective description provides an accurate approximation to the system dynamics when the period is much shorter than the typical system time-scales \cite{AdvPhys64.139(2015)}. This method has been applied in several contexts, such as the creation of synthetic gauge fields~\cite{struck2012, NaturePhoton6.782(2012), Optica2.635(2015)}, of topological insulators \cite{cayssol2013floquet,szameit_photonic_topo_insulators}, or in order to tune the coupling coefficients of quantum annealers~\cite{Floquet_annealers}. Discrete-time quantum walks can also be seen as periodically driven systems and analyzed using Floquet theory \cite{FloquetDTQW_Asboth,floquetDTQW_localization, floquetDTQW_insulators}.

In this article, we consider two central applications of Floquet engineering in the context of simulating CTQWs. The first is to use this method to simulate \emph{chiral} quantum walks, i.e., quantum walks breaking time-reversal symmetry, such as the ones presented in Ref.~\cite{ScienRep3.2361(2013)}. We provide an explicit protocol to simulate the quantum walk on a switch, where a complex coupling parameter can be used to control the propagation direction of the walk at the intersection between three branches. This protocol is easily generalized to implementing a chain of triangles, where end-to-end transport can be enhanced or suppressed by controlling the phase of the complex coupling parameters~\cite{ScienRep3.2361(2013)}. The protocol involves fast periodic driving of on-site energies, akin to the methods for creating synthetic gauge fields in optical lattices via "lattice shaking" \cite{struck2012}.  

The second application is to effectively increase the connectivity of the graph where the quantum walk takes place. Starting from a time-dependent quantum walk on a given graph, we demonstrate that it is possible to create effective next-to-nearest neighbor (NNN) couplings via suitable time-periodic modulation of the couplings of the original graph. We present two illustrative examples of this idea. In the first one, we demonstrate how to simulate NNN couplings in a one-dimensional quantum walk by periodically modulating the strength of nearest-neighbor (NN) couplings. Interestingly, our protocol introduces purely imaginary NNN couplings, which leads to time-reversal symmetry breaking and a significantly different dynamics from the usual one-dimensional CTQW.  In the second example, we consider a time-dependent quantum walk on the star graph with $N+1$ nodes and show that, although this graph has only $N$ edges, it can be used to simulate quantum walks on highly connected graphs with $\mathcal{O} (N^2)$ edges.  
Throughout the work, we discuss the possibility of implementing some of our proposals using integrated photonic circuits~\cite{szameit_review,flamini2018review}. Experiments with femtosecond laser-written waveguides have provided so far a high level of control, stability, and accuracy, making them one of the leading experimental platforms for implementing quantum walks. 

Our work is structured as follows. We start with a small introduction to Floquet theory and the Magnus expansion in Sec.~\ref{sec:theory}, explaining
how quantum walks with periodically driven Hamiltonians can be described effectively by quantum walks in a network that can be different from the original one. We derive conditions on the modulation period which guarantee that the effective description is a good approximation of the real dynamics. We then focus on engineering chiral quantum walks in Sec.~\ref{sec:ChiralWalks} from quantum walks with real coupling coefficients and periodically modulated on-site energies, covering different examples. In Sec.~\ref{sec:NNN}, we describe how NNN couplings can be introduced in different graphs via fast periodic modulation of the coupling coefficients. Finally, we present some concluding remarks in Sec.~\ref{sec:conclusions}.

\section{Floquet engineering of continuous-time quantum walks \label{sec:theory}}
We begin by considering a general model describing a quantum particle hopping in an arbitrary network. The network can be represented by a graph $G$ with $N$ sites corresponding to quantum states $\px{\ket{1}, ..., \ket{N}}$, and a set of edges $E_G$, containing the coupled pairs of sites $(i,j)$. We write the Hamiltonian describing the dynamics of the system as
\begin{align}\label{eq:QWham}
    H =\sum_i \epsilon_i \ket{i}\bra{i}+ \!\!\! \sum_{i\neq j: (i,j) \in E_G } \!\!\! J_{ij} \ket{i}\bra{j}, 
\end{align}
where the coefficients $J_{ij}$ represent the coupling strength between sites $\ket{i}$ and $\ket{j}$, with $J_{ij}=J^*_{ji}$, whereas the real parameters $\epsilon_i$ are usually interpreted as on-site energies. Mathematically, this Hamiltonian can be interpreted as the adjacency matrix of a weighted graph $G$ with edge weights $J_{ij}$ and self-loops $\epsilon_i$.

In an experimental setting, the values of $J_{ij}$ are usually real-valued tunable parameters that rapidly decay with the distance between sites. For example, in continuous-time photonic quantum walks on waveguide lattices, the evanescent coupling between waveguides decreases exponentially with the distance, and usually, only nearest-neighbor couplings are considered relevant \cite{szameit2010review}. This constraint highly restricts the kind of quantum walks that can be realized experimentally. 
However, in various experimental platforms, including photonic waveguides, the parameters describing the Hamiltonian are highly tunable and can even be made time or space-dependent. The ability to modulate in time the parameters describing the quantum walk can be exploited to simulate other types of walks that, a priori, would not be possible to implement directly due to experimental restrictions. In particular, we will focus on the case where both the on-site potentials and couplings can be tuned \emph{periodically in time} and use Floquet theory \cite{AdvPhys64.139(2015)} to derive an effective time-independent Hamiltonian describing the quantum walk we would like to engineer.

\subsection{Effective Hamiltonian and the Magnus expansion}
Let us assume we can prepare a quantum system described by a time-dependent Hamiltonian $H(t)$ of the form of Eq.~\eqref{eq:QWham} whose parameters $\epsilon_i(t)$ and $J_{ij}(t)$ are periodic functions of time with period $T$ such that $H(t+T)=H(t)$. The time-evolution operator after a period $T$ is given by
\begin{align}
   U(T)&= \mathcal{T}\left\{ e^{-\im \int_{t_0}^{t_0+T}dt~H(t)}\right\} \equiv e^{-\im H_{\text{eff}} T}
   \label{eq:effHam},  
\end{align}
where $\mathcal{T}$ represents the time-ordering operator. For simplicity, we set $t_0 = 0$ and consider units where $\hbar= 1$. The stroboscopic dynamics for times that are multiples of the period $T$ is fully described by the effective Hamiltonian $H_\text{eff}$ since 
\begin{align}
    U(m T)= U^m(T)= e^{-\im H_{\eff} ~m T}
\end{align}
for any positive integer $m$. When the frequency $\Omega=2\pi/T$ is sufficiently large, the effective Hamiltonian can be approximately computed via the Magnus expansion \cite{magnus1954} 
\begin{align}
    H_{\eff} = \sum_{n=0}^{\infty} H_{\eff}^{(n)},  
\end{align}
where each term $H_{\eff}^{(n)} = \mathcal{O}(\Omega^{- n})$. Throughout this work, we will consider at most the first two terms of this expansion, given by  
\begin{align}
H_{\eff}^{(0)} &=\frac{1}{T}\int_{0}^{T} dt \, H(t)=H_0 \label{eq:Heff0},\\
H_{\eff}^{(1)} &= \frac{1}{2 T \im } \int_0^T\int_0^{t_1} dt_1 dt_2 \, [H(t_1),H(t_2)] \label{eq:Heff1}\\
&=\frac{1}{\Omega}\sum_{l=1}^{\infty}\frac{1}{l}\left([H_l,H_{-l}]- [H_l,H_0]+[H_{-l},H_0]\right),\label{eq:Heff1fourier} \end{align}
where we have used the expansion of $H(t)$ in its Fourier series as 
\begin{eqnarray}
H(t)=\sum_{l\in\mathbb{Z}} H_l ~e^{\im l \Omega t}.
\end{eqnarray}
The idea behind our work is to simulate a target quantum walk Hamiltonian
\begin{align}\label{eq:QWham_eff}
    H_{\eff}\approx \sum_i \epsilon^{\eff}_i \ket{i}\bra{i}+  \!\!\! \sum_{i,j: (i,j)\in E_{G'} } \!\!\! J^{\eff}_{ij} \ket{i}\bra{j}.
\end{align}
by appropriately choosing the periodic functions $\epsilon_i(t)$ and $J_{ij}(t)$. In general, the effective Hamiltonian represents a quantum walk on a graph $G'$ which might be different from the original graph $G$, as we will see in Sec.~\ref{sec:NNN}. Moreover, the effective couplings can now be complex numbers, even if the couplings of the original time-dependent Hamiltonian were real. This can be used to direct the propagation of the walk and increase transport efficiency between certain nodes as we will discuss in Sec.~\ref{sec:ChiralWalks}. 

\subsection{Simulation errors due to truncation of the Magnus expansion}
The approximation of the effective Hamiltonian obtained by  truncating the Magnus expansion naturally leads to an error in the prediction of how the quantum system evolves in time. Even though this error vanishes as the period of modulation tends to zero, in reality this period will always be finite. It is thus useful to have an understanding of how large this error can be for a finite modulation period $T$ and a total evolution time $T_{\text{evol.}}=m T$, where $m$ is some positive integer. In this work, we consider the approximation $H_{\eff}\approx H_{\eff}^{(0)}$ in Sec.~\ref{sec:ChiralWalks}, whereas in Sec.~\ref{sec:NNN} we consider $H_{\eff}\approx H_{\eff}^{(0)}+ H_{\eff}^{(1)}$. Hence, we derive here the upper bounds for the simulation error in these two cases. Our aim is that, by finding this upper bound for the error, it is possible to obtain an estimate for the period $T$ that guarantees that the effective picture will provide an accurate description of the real dynamics of the system.

In order to estimate the simulation error, it is useful to consider the expansion of the unitary $U(T)$ from Eq.~\eqref{eq:effHam} as a Dyson series
\begin{align}
    U(T)=\sum_{n=0}^{\infty} \mathcal{D}_n, 
\end{align}
where 
\begin{align}
   \mathcal{D}_n= \frac{(-\im)^n}{n!}\int_0^T dt_1\dots \int_0^Tdt_n \mathcal{T} H(t_1)\dots H(t_n). 
\end{align}
We define 
\begin{align}
    H_{\text{max}}=\max_{t\in[t_0,t_0+T]} ||H(t)||
\end{align}
and assume that $T$ is small enough so that $H_{\text{max}} T < 1$. This ensures that the norm of the higher order terms of this series will become increasingly small. In fact, an upper bound for the error in the truncation of the Dyson series \cite{truncatedDyson} is given by 
\begin{align}\label{eq:error_truncDyson}
    ||U(T)- \sum_{n=0}^{K} \mathcal{D}_n||\leq \mathcal{O} \left(\frac{(H_\text{max} T)^{K+1}}{(K+1)!} \right).
\end{align}
A simple way to understand the error caused by truncating the Magnus expansion is to compare the time-evolution generated by the approximate effective Hamiltonian to the truncated Dyson series. To do so, we note that, from equations Eqs.~\eqref{eq:Heff0} and~\eqref{eq:Heff1}, we have %
\begin{align}
||H_{\eff}^{(0)}|| \, T &=\mathcal{O}(H_{\text{max}} T), \label{eq:Heff0bound}\\
||H_{\eff}^{(1)}|| \, T &= \mathcal{O}(H_{\text{max}}^2 T^2) \label{eq:Heff1bound}.      
\end{align}
We first calculate the error obtained by using the approximation $H_{\eff}\approx H^{(0)}_{\eff}$. Expanding the approximate time-evolution operator in Taylor series we obtain
\begin{align}
e^{-\im H_{\eff}^{(0)} T }\approx\mathbf{1}-\im \int_{0}^{T} dt \, H(t)+ \mathcal{O}(H_{\text{max}}^2 T^2), 
\end{align}
which is equal to the truncated Dyson series to 1st order if we neglect terms of $\mathcal{O}(H_{\text{max}}^2 T^2)$ . From Eq.~\eqref{eq:error_truncDyson} this implies that the approximation of $H_{\eff}$ by the first term in the Magnus expression leads to an error per period $T$ given by 
\begin{align}\label{eq:error0order}
    ||U(T)-e^{-\im H_{\eff}^{(0)} T }||= \mathcal{O} (H_{\text{max}}^2 T^2).
\end{align}
In order to ensure that the error is bounded after an evolution time $T_\text{evol}=mT$ we require that 
\begin{align}
    ||U( m T)-e^{-\im H_{\eff}^{(0)} m T }||\leq \epsilon\\
    \Rightarrow T=\mathcal{O}\left( \frac{\epsilon}{T_{\text{evol}}\, H_{\text{max}}^2}\right).
    \label{eq:Tbound0thorder}
\end{align}
On the other hand, if we consider the two first terms of the Magnus expansion, we can write the approximate time-evolution operator as 
\begin{align}
 e^{-\im (H_{\eff}^{(0)} + H_{\eff}^{(1)})T }
&\approx \mathbf{1}-\im (H_{\eff}^{(0)} + H_{\eff}^{(1)})T - \frac{1}{2}(H_{\eff}^{(0)}T)^2+  \mathcal{O}(H_{\text{max}}^3 T^3) \nonumber\\
&\approx \mathbf{1}-\im \int_{0}^{T} dt \, H(t)- \int_0^T \int_0^{t_1}dt_1 dt_2 \, H(t_1)H(t_2) \nonumber \\
&\quad +  \mathcal{O}(H_{\text{max}}^3 T^3),
\end{align}
which is equal to the truncated Dyson series at 2nd order, if we neglect terms of $\mathcal{O}(H_{\text{max}}^3 T^3)$. Hence, via Eq.~\eqref{eq:error_truncDyson} we obtain a bound on the error per period
\begin{align}\label{eq:error1order}
    ||U(T)-e^{-\im (H_{\eff}^{(0)}+H_{\eff}^{(1)}) T }||=\mathcal{O}(H_{\text{max}}^3 T^3).
\end{align}
Hence, the total error after a time-evolution $T_\text{evol}=mT$ can be bounded by $\epsilon$, by choosing the period $T$ such that
\begin{align}\label{eq:Tbound1storder}
    T=\mathcal{O}\left(\frac{\sqrt{\epsilon}}{\sqrt{T_\text{evol}} (H_\text{max})^{3/2}}\right).
\end{align}
We can thus use Eqs.~\eqref{eq:Tbound0thorder} and~\eqref{eq:Tbound1storder} to estimate the appropriate modulation period $T$, depending on the system energy scale $H_{\text{max}}$ and the desired total evolution time. Note, however, that these upper bounds are not necessarily tight and it is possible that in specific cases the error in Eqs.~\eqref{eq:error0order} and \eqref{eq:error1order} is overestimated. For this reason, we also provide throughout this work numerical simulations of the real dynamics in order to have a more accurate comparison between the real and effective models. Nevertheless, the bounds we provide are of independent interest and can be particularly useful when the real dynamics of the quantum systems involved is hard to compute numerically. 

\section{Simulation of quantum walks breaking time-reversal symmetry \label{sec:ChiralWalks}}
In  Ref.~\cite{ScienRep3.2361(2013)}, it was shown that complex coupling coefficients in a CTQW quantum walk can be used to break time-reversal symmetry and control the direction of propagation of the particle. This leads to an increase in transport efficiency between an initial and a target node in certain graphs. However, in an experimental setting, tunable complex coupling coefficients are not always readily available. 
Here, we discuss how these effects can be simulated via quantum walks with \emph{real} coupling coefficients and \emph{periodically modulated} on-site potentials and briefly discuss the possibility to implement our simulation protocol in integrated photonics circuits, one of the leading experimental platforms for implementing photonic quantum walks.

\subsection{Time-reversal symmetry breaking in quantum walks}
Provided that the coupling parameters of a Hamiltonian are real, the dynamics of the system respects probability time-reversal symmetry (TRS) \cite{chiralQW} in the following sense: the transition probability between two states $\ket{i}$ and $\ket{j}$ after forward time evolution 
\begin{align}
    P(t)_{i\rightarrow j}= |\bra{j}e^{-\im H t}\ket{i}|^2
\end{align}
equals the transition probability of the time-reversed process 
\begin{align}
    P(-t)_{i \rightarrow j} &= |\bra{j}e^{\im H t}\ket{i}|^2 =|\bra{j}e^{- \im H^* t}\ket{i}|^2,  
\end{align}
since $H = H^*$. Equivalently, this symmetry can be interpreted as the absence of a directional bias in the propagation as 
\begin{align}
   & P(t)_{i \rightarrow j}=P(-t)_{i \rightarrow j} 
    \Leftrightarrow  P(t)_{i \rightarrow j}=P(t)_{j \rightarrow i}.
\end{align}
Thus, a necessary condition to break probability TRS in quantum walks is to have complex coupling coefficients. However, this alone is not sufficient. The transition probabilities are invariant under gauge transformations $\ket{i}\rightarrow e^{\im \phi_i}\ket{i}$, which induce a transformation of the coupling parameters $J_{ij}\rightarrow e^{\im(\phi_i-\phi_j)}J_{ij}$. If there is a gauge transformation such that the transformed couplings are real, then probability TRS is still respected. 

More general conditions which ensure probability TRS are derived in Ref.~\cite{chiralQW}, wherein the authors show that for structures such as trees (which includes the linear chain) and bipartite graphs this symmetry is always respected. For graphs that do not fall in these categories, the presence of complex coupling coefficients can lead to probability TRS breaking. 

Throughout the work, we refer to probability TRS breaking simply as TRS symmetry breaking. In Ref.~\cite{chiralQW}, the authors also discuss the concept of amplitude TRS breaking i.e., when $\bra{j}e^{\im H t}\ket{i}\neq\bra{j}e^{- \im H t}\ket{i}$. Since this does not necessarily lead to directional bias of transition probabilities, in our work we focus only on probability TRS breaking. 

\subsection{Simulating complex couplings via Floquet engineering}
In order to show how to simulate quantum walks breaking TRS, we use a technique known as time-asymmetric lattice shaking~\cite{struck2012}, introduced in the context of experiments with ultracold atoms with the aim of creating synthetic magnetic fields for neutral particles.

Let us consider the time-dependent periodic Hamiltonian
\begin{align}\label{eq:QWham_TD}
    H(t)=\sum_i \beta_i(t) \ket{i}\bra{i}+ \!\!\! \sum_{i,j: (i,j)\in E_G } \!\!\! J_{ij} \ket{i}\bra{j}, 
\end{align}
where the on-site terms $\beta_i(t)$ are periodic functions with period $T$. To analyse the dynamics of the system it is useful to consider the rotated basis $\ket{\psi'(t)}=V(t)\ket{\psi(t)}$, where we define  
\begin{align}
    V(t)= \exp\left(\im \sum_i \mathcal{V}_i (t)\ket{i} \bra{i}\right)
\end{align}
and
\begin{align}
\mathcal{V}_i (t) = \int_{t_0}^t dt' \, \beta_i (t') - \av{\int_{t_0}^t dt' \, \beta_i (t')}_T.
\label{eq:Widef}
\end{align}
Here, we denote the time average of a function $f(t)$ over a period $T$ as  $\av{f(t)}_T=T^{-1}\int_{t_0}^{t_0+T}\!\!\!\!\!f(t) ~dt$. Note that the outcome probabilities of a measurement of either $\ket{\psi(t)}$ or $\ket{\psi'(t)}$ in the site basis are the same so, for simplicity, in what follows we will consider only $\ket{\psi'(t)}$. 

The wavefunction $\ket{\psi'(t)}$ obeys the Schr\"{o}dinger equation with Hamiltonian 
\begin{align}
    \hat{H}'(t)&=V(t) \, \hat{H}(t)\, V^{\dagger}(t) + \im \dot{V}(t) \, V^\dagger(t)\\
        &= \sum_{\av{ij}} J_{ij} \, \exp \pr{\im \pc{\mathcal{V}_i (t) -\mathcal{V}_j(t)} }\ket{i} \bra{j}.
\end{align}
We assume the period $T$ is small enough, so that the effective Hamiltonian is well approximated by the first term of the Magnus expansion
\begin{eqnarray}
H_\text{eff} \approx \av{H'(t)}_T =  \sum_{\av{ij}} J_{ij}^\text{eff} \ket{i}\bra{j},
\label{eq:Heff}
\end{eqnarray}
with the renormalized coupling terms
\begin{align}
J^\text{eff}_{ij} = J_{ij} \av{\exp \pr{\im \mathcal{V}_{ij} (t)}}_T,  
\label{eq:Jeff}
\end{align}
where we have defined $\mathcal{V}_{ij} = \mathcal{V}_i -\mathcal{V}_j$.

The imaginary part of the effective tunneling in Eq.~\eqref{eq:Jeff}
\begin{align}
\text{Im} \pc{\frac{J^\text{eff}_{ij}}{J_{ij}}} &= \av{\sin \pc{\mathcal{V}_{ij}(t)}}_T \label{eq:ImJeff}
\end{align}
can only be non-zero provided that the function $\mathcal{\beta}_{ij} (t)=\beta_i(t)-\beta_j(t)$ breaks two fundamental symmetries~\cite{struck2012}: (i) The function $\mathcal{\beta}_{ij} (t)$ needs to break inversion symmetry
\begin{eqnarray}
\mathcal{\beta}_{ij} (t-\tau) = - \mathcal{\beta}_{ij} (-t-\tau) \label{eq:invsymmetry}
\end{eqnarray}
with respect to all points $\tau$ in time, and (ii) The function $\mathcal{\beta}_{ij} (t)$ needs to break shift inversion symmetry
\begin{eqnarray}
\mathcal{\beta}_{ij} (t) = -\mathcal{\beta}_{ij} \pc{t-T/2}. \label{eq:shiftinversionsymmetry}
\end{eqnarray}
The specific form of the functions $\beta_i(t)$ depends on the effective quantum walk Hamiltonian we would like to simulate. In what follows we present a specific protocol to simulate the simplest of the Hamiltonians which leads to time-reversal symmetry (a triangular loop). In addition, we show how to use this protocol to engineer some of the quantum walk Hamiltonians considered in Ref.~\cite{ScienRep3.2361(2013)}, where it is possible direct the walker and increase the transport efficiency between certain nodes using effective complex phases.

\subsubsection{Triangular loop}\label{subsec:triangle}
\begin{figure}
\centering
\includegraphics{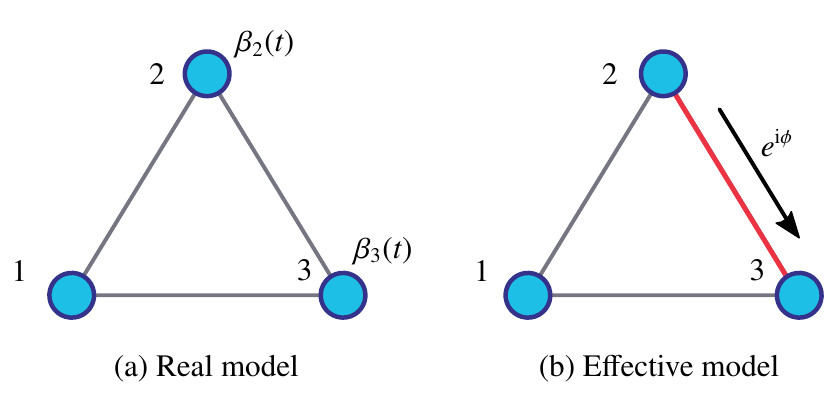}
\caption{(Color online) a) Scheme representing a quantum walk on a triangle where the site-energies of nodes $2$ and $3$ are modulated in time with functions $\beta_2(t)$ and $\beta_3(t)$. b) The model represented in a) can be used to simulate the dynamics of a quantum walk on a triangle with a complex coupling coefficient between nodes $2$ and $3$ represented by the red edge.This is the simplest quantum walk breaking time-reversal symmetry and can be used as a building block in more complex structures in order to direct quantum transport.}
\label{fig:triangle}
\end{figure}
The crucial building block for some of the Hamiltonians discussed in Ref.~\cite{ScienRep3.2361(2013)} is a triangle loop where the strength of the couplings is uniform $|J_{ij}|=J$ but one of the couplings, say between nodes 2 and 3, is complex with $J_{23}=J e^{\im \phi}$ (see Fig.~\ref{fig:triangle}~b). Note that this Hamiltonian is equivalent, up to a gauge transformation, to a Hamiltonian where 
\begin{align}\label{eq:couplings_triangle}
J_{12}=J_{23}=J_{31}=J e^{\im \phi/3}. 
\end{align}
The protocol we consider for the simulation of this system involves a fast time-periodic modulation of two of the on-site potentials, say $\beta_2(t)$ and $\beta_3(t)$, while keeping $\beta_1(t) = 0$, (see Fig.~\ref{fig:triangle}~a) ). Namely, we consider the Hamiltonian
\begin{equation}\label{eq:QWham_triangle}
    H(t)= \beta_2(t) \ket{2}\bra{2}+\beta_3(t) \ket{3}\bra{3}+ \sum_{i,j: (i,j)\in E_G } J' \ket{i}\bra{j}, 
\end{equation}
where in this case the graph $G$ is a triangle with uniform real couplings given by $J'$.

A possible form for these functions, which leads to complex coupling coefficients,  is based on sine functions that are turned on and off for a certain time, as proposed in Refs.~\cite{struck2012, NatPhys9.738(2013)} in the context of simulating artificial magnetic fields for ultracold atoms in triangular lattices.  Here, we consider an alternative form for the modulation based on step functions, motivated by the possibility to implement our scheme in integrated photonic circuits. The reason is twofold. Firstly, we show that these step-like modulation functions create higher effective complex phases for the same modulation amplitude, when compared to the sine-like modulation of Ref.~\cite{struck2012} (see more details for the results using sine-function modulation in Appendix~\ref{app:alternative_modulation}). This can be important if in an experimental setting the maximum modulation amplitude is restricted. Secondly, step-like variations of waveguides' propagation constant have already been experimentally demonstrated in integrated photonic circuits, as it will be discussed in Sec.~\ref{sec:experimentscomplexphases}.

\begin{figure}
    \centering
    \includegraphics{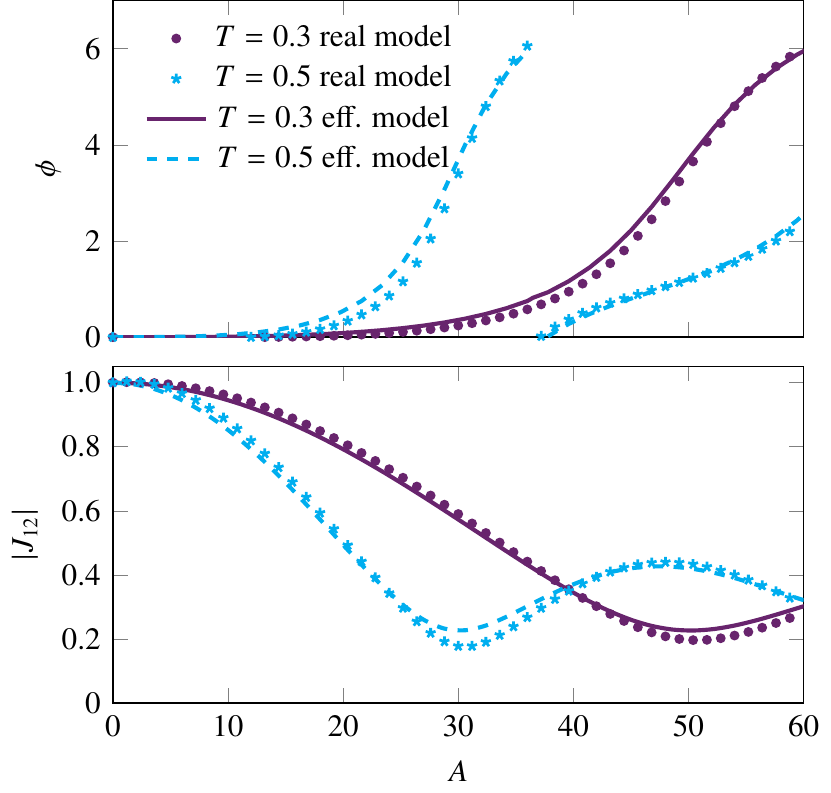}
    \caption{(Color online) Comparison between real (marks) and effective models (lines) for the accumulated phase around the triangular loop $\phi$ from Eq.~\eqref{eq:eff_coupling_step} (top) and absolute value of coupling $|J_{12}|$ (bottom) for two values of the period $T=0.3$ (in purple) and 0.5 (in blue) as a function of the modulation amplitude $A$, for step-like modulation functions. }
    \label{fig:comparison_realvseff_step}
\end{figure}

The step-like modulation functions we consider are
 \begin{align}
\beta^{(\text{step})}_2(t)&=\begin{cases}A ,~0\leq t \leq \frac{T}{3},\\~\\
 - A,  ~ \frac{T}{3} <t\leq \frac{2T}{3}\\~\\0,  ~ \frac{2T}{3} <t\leq T,\end{cases}\label{eq:beta2step}
 \end{align}
 \begin{align}
 \beta^{(\text{step})}_3(t)&=\begin{cases}0 ,~0\leq t \leq \frac{T}{3},\\~\\
 - A,  ~ \frac{T}{3} <t\leq \frac{2T}{3}\\~\\A,  ~ \frac{2T}{3} <t\leq T.\end{cases}\label{eq:beta3step}
\end{align}
From Eq.~\eqref{eq:Jeff}, the effective couplings obtained are given by
\begin{align}
J^{\text{eff}}&=J_{12}^{\text{eff}}=J_{23}^{\text{eff}}=J_{31}^{\text{eff}}\nonumber \\
&=J' e^{\im \frac{AT}{9}}\left[\frac{1}{3}+\frac{2 \im}{AT}\left(e^{-\im \frac{AT}{3}}-1\right)\right]\label{eq:eff_coupling_step}.
\end{align}   
Hence, this choice of the modulation functions ensures that both the absolute value of all couplings as well as their phase is the same along the loop $1\rightarrow 2 \rightarrow 3\rightarrow 1$ as in Eq.~\eqref{eq:couplings_triangle}.
The value of the effective phase is then given by 
\begin{align}\label{eq:eff_phase}
\phi =3 \text{Arg}(J^{\text{eff}})~~\text{mod}~2\pi.
\end{align} 
In Fig.~\ref{fig:comparison_realvseff_step}, we compare the parameters $|J^\text{eff}|$ and $\phi$, extracted from a numerical calculation of $H_{\eff}=\im \log(U(T))/T$ and from the analytical calculation from Eq.~\eqref{eq:eff_coupling_step}, showing a good agreement between the two for the range of parameters shown. Note that the effective coupling $|J^\text{eff}|$ is lower than the value of $J'$. Hence, in order to achieve a desired strength $J$ for the effective coupling, the value of $J'$ has to be chosen accordingly, taking into account the modulation amplitude $A$ and the period $T$.
%
\subsubsection{Switch and chain of triangles}\label{sec:other_structures}

\begin{figure}
\centering
\includegraphics{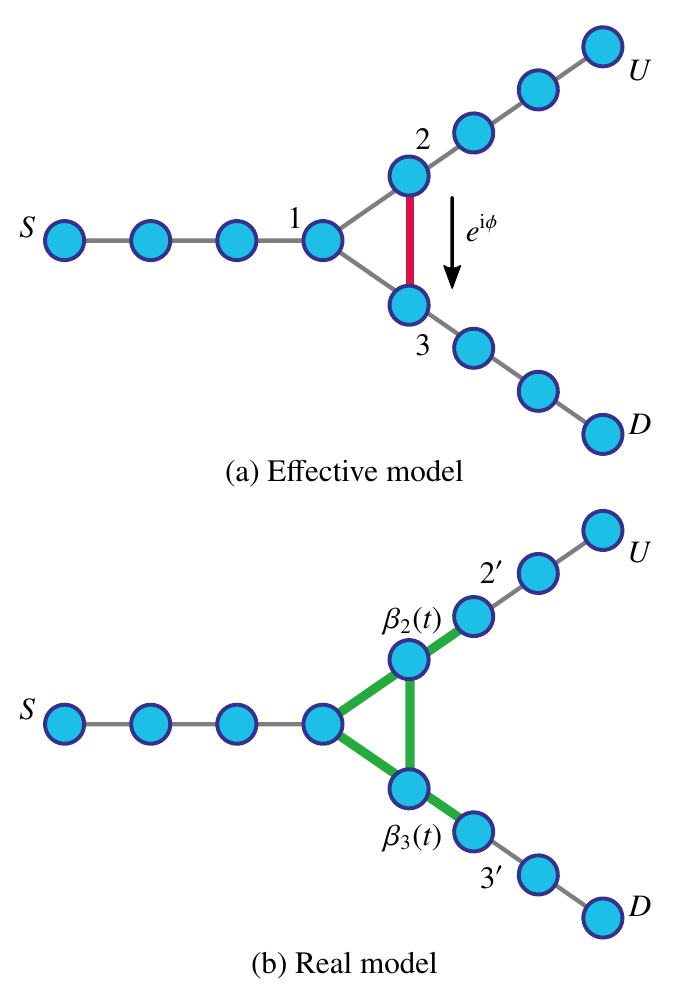}
\caption{(Color online) Scheme of the switch structure. a) The presence of a complex-valued coupling (red link) in the central triangle allows the biasing of the propagation direction of quantum walk starting at $S$ towards $U$ or towards $D$. This effect can be simulated via a quantum walk with \emph{real} coupling coefficients and periodically modulated site-energies at nodes 2 and 3, with modulation functions $\beta_2(t)$ and $\beta_3(t)$ (see Eqs.~\eqref{eq:beta2step} and \eqref{eq:beta3step}). The strength of couplings inside the triangular loop as well as those between 2 and 2' and nodes 3 and 3' (represented by the thicker green edges), need to be tuned appropriately in order to avoid backscattering.} 
\label{fig:switch}
\end{figure}

In Ref.~\cite{ScienRep3.2361(2013)}, the simplest example that was constructed to illustrate the role of complex coupling coefficients in a quantum walk was the switch represented in Fig.~\ref{fig:switch}~a). 
Starting from node $S$, the direction of the propagation of the quantum walk can be controlled by the complex hopping phase present in the central triangular loop. To do so, the Hamiltonian of the system is chosen such that all coupling coefficients are real, with strength $J=1$ (which sets our energy units), except for the coupling between nodes 2 and 3, which is set to $J_{23}=e^{\im \phi}$. If the phase $\phi$ is zero, due to the graph's reflection symmetry, the probability of observing the particle at either nodes $U$ or $D$ is the same. A non-zero value of $\phi$ breaks this reflection symmetry (and time-reversal symmetry) and biases the direction of propagation. It was observed in Ref.~\cite{ScienRep3.2361(2013)} that the maximum bias is achieved for values of $\phi=\pi/2$ (towards $U$) or $\phi=3\pi/2$ (towards $D$). 

To simulate this quantum walk, we choose a time-dependent Hamiltonian with real parameters and periodically-modulated site-energies at nodes 2 and 3, as represented in Fig.~\ref{fig:switch}~b).
We choose the step-like modulation functions from Eqs.~\eqref{eq:beta2step} and \eqref{eq:beta3step}, but similar results could be obtained for the sine-like modulations discussed in Appendix~\ref{app:alternative_modulation}. The modulation amplitude, as well as the period, can be tuned following two requirements: the value of $A T$ is chosen to generate the desired effective phase, following Eqs.~\eqref{eq:eff_coupling_step} and \eqref{eq:eff_phase};  the period $T$ is chosen to be short enough, so the effective model provides an accurate description of the real dynamics. The latter requirement can be fulfilled by using the bound from Eq.~\eqref{eq:Tbound0thorder}. However, for this particular example, we have numerically found that this constraint is too conservative, and higher values of $T$ still guarantee a good correspondence between the dynamics of the real model and the effective one. 

\begin{figure}
\centering
\includegraphics{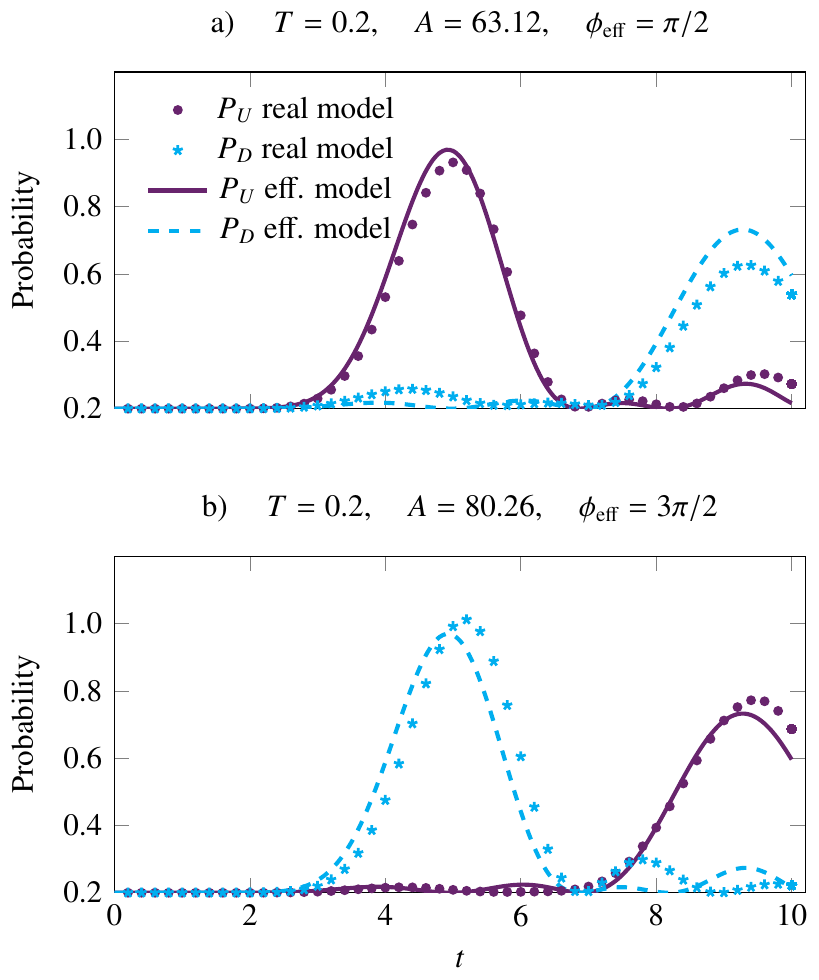}
\caption{(Color online) Comparison between the probabilities of observing the particle at nodes $U$ and  $D$ for the quantum walk on the switch, starting from node $S$ (see Fig.~\ref{fig:switch}). The lines correspond to the dynamics of effective model with a complex phase $\text{Arg}(J_{23})=\phi_{\eff}=\pi/2$  in (a) and $\text{Arg}(J_{23})=\phi_{\eff}=3 \pi/2$ in (b). The marks correspond to the stroboscopic dynamics of the real model, where the on-site energies of nodes $2$ and $3$ are modulated periodically according to the step-like functions from Eqs.~\eqref{eq:beta2step} and \eqref{eq:beta3step}. The modulation period was chosen to be $T=0.2$ and the modulation amplitude $A=63.12$ in (a) and $A= 80.2575$ in (b) so that the real dynamics approximately simulates the quantum walk on the switch with uniform coupling strength and a complex hopping phase $\text{Arg}(J_{23})=\pi/2$ and $\text{Arg}(J_{23})=3 \pi/2$, respectively.  Our units are chosen such that the effective Hamiltonian has uniform couplings  $|J^{\eff}|=1.$\label{fig:probs_switch}}
\end{figure}

The results are summarized in Fig.~\ref{fig:probs_switch}, wherein we compare the probabilities of observing the particle at $D$ and $U$, computed via the real (represented by the marks) and the effective model (represented by the continuous and dashed lines). We chose the initial state of the quantum walk to be localized at node $S$, the modulation period $T=0.2$, and two values of modulation amplitude in order to create an effective phase of $\pi/2$ in Fig.~\ref{fig:probs_switch}~a) and of $3\pi/2$ in Fig.~\ref{fig:probs_switch}~b). We observe a good agreement between the system's real and effective descriptions, especially for lower evolution times. The periodic modulation of the site-energies at nodes $2$ and $3$ of the central triangle leads to a significant bias in the direction of propagation of the wavefunction. For an effective phase $\phi_{\eff}=\pi/2$, the particle goes mostly towards $U$, reaching a maximum probability $\sim 0.8$ at this site, whereas for $\phi_{\eff}=3 \pi/2$ a similar bias is achieved towards $D$, in agreement with Ref.~\cite{ScienRep3.2361(2013)}. 

Note that in order to ensure that all the couplings of $H_{\eff.}$ have a uniform strength $|J_{\eff.}|=1$ (in appropriate units), the values of the coupling strength of the time-dependent Hamiltonian $H(t)$ have to be chosen appropriately. Namely, the couplings represented in green in Fig.~\ref{fig:switch}~b), which represent connections to either node $2$ or $3$, need to be stronger than the rest in order to compensate for the weakening of the effective coupling strength due to the periodic site-energy modulation of these nodes, which is visible in Fig.~\ref{fig:comparison_realvseff_step}~b). If such compensation is not done, we observe via numerical simulations that there is a significant backscattering of the particle at the central triangle, washing out the sharp probability peaks at nodes $U$ and $D$ observed in Fig.~\ref{fig:probs_switch}~a) and b), respectively. 

\begin{figure}
\centering
\includegraphics{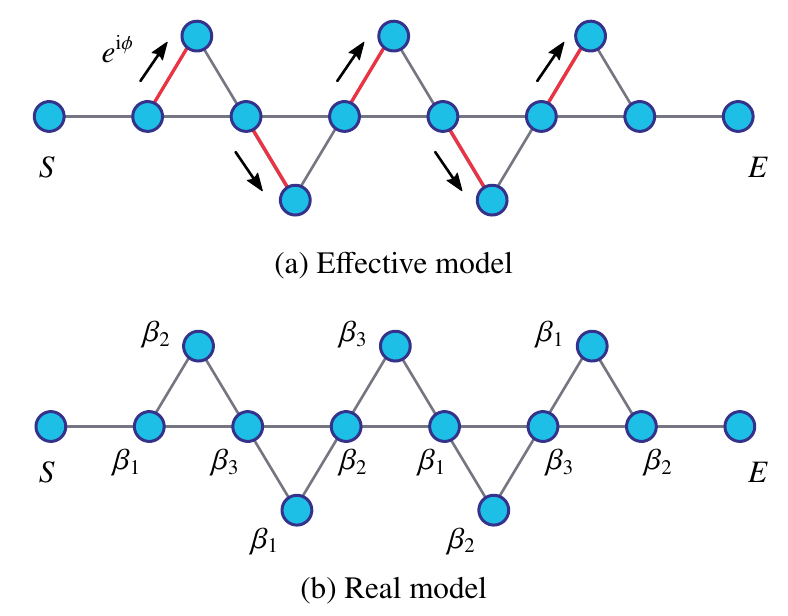}
\caption{(Color online) Chain of coupled triangles. a) In the effective model the complex hopping phase $e^{\im \phi}$ represented by the red links can be used to enhance the transport efficiency between nodes $S$ and $E$ \cite{ScienRep3.2361(2013)}. These complex phases can be effectively implemented by modulating the site energies via the periodic functions $\beta_1(t)$ (which can be chosen to be $\beta_1(t)=0$), $\beta_2(t)$ and $\beta_3(t)$, according to the pattern represented in panel b).}
\label{fig:chain_triangles}
\end{figure}

We remark that the protocol to create complex hopping phases in a triangle from Sec.~~\ref{subsec:triangle} can be readily used to simulate the quantum walks in the structure represented in Fig.~\ref{fig:chain_triangles}~a), also studied in Ref.~\cite{ScienRep3.2361(2013)}. This structure forms a chain of coupled triangular loops, where in each loop one of the coupling coefficients has a complex phase $\phi$ as in Fig.~\ref{fig:triangle}~a). It has been shown in Ref.~\cite{ScienRep3.2361(2013)} that these complex phases can either significantly enhance or suppress the transport efficiency between the two ends of the chain. In order to simulate this quantum walk  on-site energies should be modulated according to the pattern shown in Fig.~\ref{fig:chain_triangles}~b), where the functions $\beta_2(t)$ and $\beta_3(t)$ are chosen according to Sec.~~\ref{subsec:triangle}.


\subsection{Experimental feasibility}\label{sec:experimentscomplexphases}
Several proposals have been put forward to simulate quantum dynamics showing time-reversal breaking for atomic ensembles \cite{struck2012}, or coherent light \cite{NaturePhoton6.782(2012), szameit_photonic_topo_insulators, Optica2.635(2015)}. These proposals are driven by the systems' desirable features, such as transport along the edge states that are robust against back-scattering. However, quantum walks with effective complex hopping phases in the single or few particle regimes have not yet been realized. Here, we argue that the proposals discussed in Sec.~\ref{sec:other_structures} could, in principle, be implemented by coupling waveguides in photonic integrated circuits \cite{flamini2018review}. This is an experimental platform that has extensively been used to implement and study quantum walks \cite{szameit_review}. 
The waveguide fabrication is done through a permanent change in the index of refraction induced by nonlinear absorption phenomena when a femtosecond laser is focused on a glass substrate. A photon can hop from a waveguide to the neighboring one via evanescent coupling, and the coupling strength (i.e., the $J_{ij}$ parameters in Eq.~\eqref{eq:QWham_TD}) can be precisely controlled by tuning the distance between the waveguides. Each waveguide represents a node of the quantum walk, and planar graphs like those represented in Fig.~\ref{fig:switch}, or Fig.~\ref{fig:chain_triangles} can be realized by exploiting laser witting in three-dimensions \cite{caruso2016fast, perez2018endurance}. Furthermore, the waveguide's propagation constant along the photon's propagation direction plays the role of a time-dependent on-site potential felt by the photon (i.e., the $\beta_i(t)$ parameters in Eq.~\eqref{eq:QWham_TD}). The latter depends in a non-trivial way on the waveguide refractive index and can be tuned by adjusting the laser intensity in the writing process \cite{JPhysB43.163001(2010)}. This technique has been used to create random time-dependent step-like potentials to simulate decoherence in photonic quantum walks \cite{caruso2016fast,perez2018endurance}. Hence, it is in principle possible to implement different time-asymmetric periodic modulation of the waveguides' propagation constants, according to the functions proposed in Sec.~\ref{subsec:triangle}.

Interestingly, since the waveguide's propagation constant is also a function of the photon wavelength, photons of different wavelengths would feel different amplitudes of the time-dependent on-site potential. The scheme we propose to simulate the quantum walk on the switch (see Fig.~\ref{fig:switch}) thus creates a wavelength-dependent effective complex hopping phase, which, in turn, translates into a wavelength-dependent direction bias for the photonic quantum walk. The proposed scheme could open the possibility to experimentally study the effect of time-reversal symmetry breaking in photonic quantum walks in the single or few-photon regime and exploit effective complex hopping phases to direct photon propagation. 

\section{Simulating next-to-nearest neighbor interactions via time-periodic Hamiltonians}\label{sec:NNN}
In physical systems, the coupling strength between two nodes of a quantum walk tends to decrease with the distance between them, thereby restricting the connectivity of the walks that can be implemented. In this section, we show that time-dependent modulation of the couplings in a quantum walk can be exploited to selectively  introduce effective couplings between NNN. In particular, we demonstrate that the relative strength between the NN and NNN couplings can be fully controlled. We illustrate this idea for the 1D walk and for the quantum walk on the star graph. The NNN couplings introduced are purely imaginary and can also be used to study time-reversal symmetry breaking effects.

\subsection{Introducing effective NNN couplings on general graphs}
We begin by describing a quantum walk on a graph $G$ with a general time-dependent Hamiltonian given by
\begin{align}\label{eq:QWham2}
    H(t)=\sum_{i,j} J_{ij}(t)\, \ket{i}\bra{j}, 
\end{align} 
where $J_{ij}(t+T)=J_{ij}(t)$ is time-periodic with real parameters and $J_{ij}(t)=J_{ji}(t)$. We assume the on-site terms to be zero. For simplicity, we consider the sum over all pairs $(i,j)$ and assume $J_{ij}(t)=0$ if $(i,j)$ is not an edge of the graph $G$. We can expand each of the coupling parameters in the Fourier series as 
\begin{align}\label{eq:couplingsFourier}
    J_{ij}(t)= J^{(0)}_{ij}+\sum_{l=1}^{\infty}J^{(l)}_{ij}\cos \pc{l \Omega t+ \phi^{(l)}_{ij}},  
\end{align}
where $J^{(l)}_{ij}=J^{(l)}_{ji}$ and $\phi^{(l)}_{ij}=\phi^{(l)}_{ji}$. In contrast to Sec.~\ref{sec:ChiralWalks}, we will consider the first two terms of the Magnus expansion, i.e. $H_{\eff}\approx H_{\eff}^{(0)}+H_{\eff}^{(1)}$. Then, from Eq.~\eqref{eq:Heff0}, we find 
\begin{align}\label{eq:H0general}
  H_{\eff}^{(0)}= \sum_{i,j} J^{(0)}_{ij} \ket{i}\bra{j} 
\end{align}
as, at zero-order in the Magnus expansion, only the time-independent part of the Fourier expansion of the Hamiltonian contributes. To calculate the 1st order term using Eq.~\eqref{eq:Heff1fourier}, we start by evaluating the following terms 
\begin{align}
\bra{i}[H_l, H_{-l}]\ket{k}&= \frac{\im}{2} \sum_{j} J^{(l)}_{ij}J^{(l)}_{jk} \sin{(\phi^{(l)}_{ij}-\phi^{(l)}_{jk})} , 
\\
 \bra{i}[H_l, H_{0}]\ket{k}&= \frac{1}{2}\sum_{j} \pr{J^{(l)}_{ij}J^{(0)}_{jk}e^{\im \phi^{(l)}_{ij}}-J^{(0)}_{ij}J^{(l)}_{jk}e^{-\im \phi^{(l)}_{jk}}},
\\
 \bra{i}[H_{-l}, H_{0}]\ket{k}&= \frac{1}{2} \sum_{j} \pr{J^{(l)}_{ij}J^{(0)}_{jk}e^{-\im \phi^{(l)}_{ij}}-J^{(0)}_{ij}J^{(l)}_{jk}e^{\im \phi^{(l)}_{jk}}}. 
\end{align}
This leads to the following expression for the matrix entries of $H_{\eff}^{(1)}$ 
\begin{align}
  \bra{i}H_{\eff}^{(1)}\ket{k}=&\sum_{l=1}^{\infty}\frac{\im}{ l \Omega}\sum_{j}  \left[ \frac{1}{2} J^{(l)}_{ij}J^{(l)}_{jk} \sin{(\phi^{(l)}_{ij}-\phi^{(l)}_{jk})} \right. \nonumber\\& \left. - J^{(l)}_{ij}J^{(0)}_{jk}\sin{\phi^{(l)}_{ij}}+ J^{(0)}_{ij}J^{(l)}_{jk} \sin{\phi^{(l)}_{jk}} \right].\label{eq:NNNcouplings} 
\end{align}
Note that $\bra{i}H_{\eff}^{(1)}\ket{i}=0$ as $\phi^{(l)}_{ij}=\phi^{(l)}_{ji}$. 
Eq.~\eqref{eq:NNNcouplings} can only be non-zero if there is a site $j$ such that both $(i,j)$ and $(j,k)$ are edges of $G$, i.e., if there exists at least one path of length 2 from $i$ to $k$.
These effective NNN interactions are purely imaginary and can be used to simulate quantum walks breaking time-reversal symmetry. Note that imaginary NNN couplings also appear from the Magnus expansion of quantum systems with periodically driven on-site energies \cite{Eckardt_2015}. However, in our case, we will see that the ability to periodically modulate the coupling coefficients gives us extra freedom to control in a simpler way the relative strength between NN and NNN couplings.

\subsection{Simulating imaginary NNN neighbor couplings in a 1D quantum walk}\label{sec:1d_imNNN}
\begin{figure}
\centering
\includegraphics{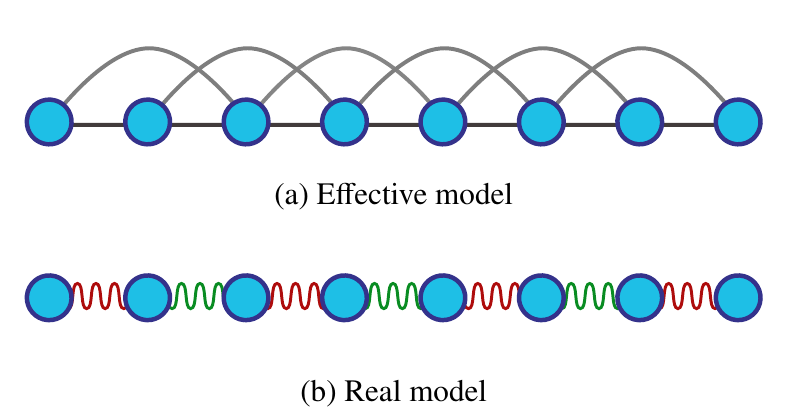}
\caption{(Color online) a) Representation of the graph corresponding to a one-dimensional quantum walk with NN and NNN couplings. b) The dynamics of this quantum walk can be simulated by a Hamiltonian with oscillating coupling coefficients with two different oscillating phases, represented by the red and green wobbly lines. The Hamiltonians corresponding to the real and effective models are shown in Table~\ref{table:realandeffHams}.  }
\end{figure}

\begin{figure}
    \centering
    \includegraphics{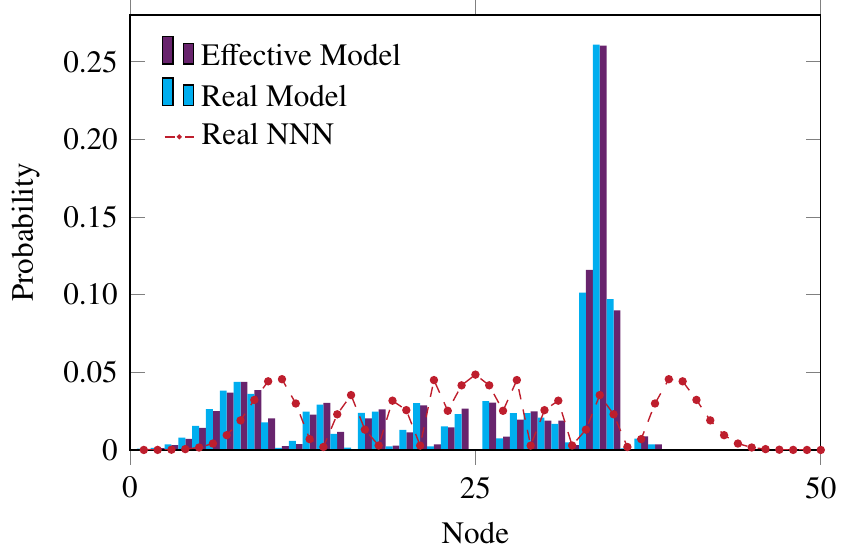}
  \caption{(Color online) The blue bars represent the probability distribution of a one-dimensional quantum walk on a line of 50 nodes with imaginary NNN couplings after time $T_\text{evol}=7$. The initial state was chosen to be localized in node 25, the NN coupling strength was set to $K_1=1$ (arbitrary units) and the imaginary NNN coupling to $K_2=0.2 i$. The purple bars represent the probability distribution of the real dynamics generated by our simulation protocol, which uses the time-dependent periodic Hamiltonian with NN couplings from Table~\ref{table}. We have used the values for the period $T=0.5$ and chosen the parameters $J^{(0)}=3 K_1 T/(4\pi)$, $J^{(1)}= \sqrt{|K_2|}$ and total evolution time of $T_\text{sim}= 4\pi T_\text{evol}/(3T)$. The red dots represent the dynamics of the quantum walk with $K_1=1$  and real NNN coupling to $K_2=0.2$.} 
    \label{fig:QW_imNNN}
\end{figure}

To simplify the expression in Eq.~\eqref{eq:NNNcouplings}, we make the following choice: apart from the time-independent term given by $J^{(0)}_{ij}$, we choose two modulation frequencies, say $\Omega$ and $2 \Omega$, such that 
\begin{align}
  \phi^{(2)}_{ij}&=\phi^{(1)}_{ij}+\pi,\\
  J^{(2)}_{ij}&=2 J^{(1)}_{ij}, 
\end{align}
while keeping $J^{(l)}_{ij}=0$ for $l>2$. This allows us to eliminate the 2nd and 3rd term of Eq.~\eqref{eq:NNNcouplings} and arrive at
\begin{align}
    \bra{i}H_{\eff}^{(1)}\ket{k}=\frac{3 \im}{ 2 \Omega} \sum_{j}  J^{(1)}_{ij}J^{(1)}_{jk} \sin{(\phi^{(1)}_{ij}-\phi^{(1)}_{jk})}. \label{eq:NNNcouplings2}
\end{align}
Let us now consider a 1D system with a time-dependent periodic Hamiltonian given by  
\begin{align}
    H(t)= \sum_{j=1}^{N-1}J_{j,j+1}(t) \; (\ket{j}\bra{j+1}+\ket{j+1}\bra{j}).
\end{align}
As a further simplification, we assume the uniformity of the terms $J^{(0)}_{j, j+1}=J^{(0)}$ and $J^{(1)}_{j, j+1}=J^{(1)}$. In this case, from Eq.~\eqref{eq:NNNcouplings2} it can be shown that the only entries of $H_{\eff}\approx H_{\eff}^{(0)}+ H_{\eff}^{(1)}$ that can be non-zero are
\begin{align}
\bra{j}H_{\eff}\ket{j+1}&= J^{(0)},\\
\bra{j}H_{\eff}\ket{j+2}&=\frac{3 \im }{2 \Omega} (J^{(1)})^2 \sin \pc{\phi^{(1)}_{j,j+1}-\phi^{(1)}_{j+1,j+2}}, \label{eq:NNNcouplings3}
\end{align}
as well as its complex conjugate terms. The strength of each of the NNN couplings can be controlled individually via the phase difference $\phi^{(1)}_{j,j+1}-\phi^{(1)}_{j+1,j+2}$. A uniform strength can be achieved with the choice 
\begin{align}
    \phi^{(1)}_{j,j+1}\equiv  - j \frac{\pi}{2}. 
\end{align}
From Eq.~\eqref{eq:NNNcouplings3} this yields
\begin{align}
\bra{j}H_{\eff}\ket{j+2}&=\frac{3 \im}{2 \Omega} (J^{(1)})^2.
\end{align}
These choices thus lead  to a protocol to simulate imaginary NNN couplings in 1D quantum walks with uniform strength, which we summarize in Table~\ref{table:realandeffHams}. As previously mentioned, the effective NNN coupling coefficients obtained via this protocol are purely imaginary and, in general, cannot be made real via a gauge transformation. This leads to observable physical consequences such as the loss of reflection symmetry of the probability distribution of a quantum walk starting in the middle of the 1D chain, which can be seen in Fig.~\ref{fig:QW_imNNN}. 

More precisely, let us define a family of 1D quantum walks with NN couplings $K_1$ and NNN couplings given by $K_2$, with Hamiltonian 
\begin{align}\label{eq:QW_NNN}
    H_{1D}(K_1,K_2)=& \sum_j K_1\ket{j}\bra{j+1}+K_1^*\ket{j+1}\bra{j})\nonumber \\&+\sum_j  K_2 \ket{j}\bra{j+2}+ K_2^* \ket{j+2}\bra{j}.
\end{align}
It can be seen that if either $K_1$ or $K_2$ are zero, the probability distribution after a certain time $t$ has a reflection symmetry around the starting node. More generally, if we write $K_1=|K_1|e^{\im \phi_1}$ and $K_2=|K_2|e^{\im \phi_2}$, it can be shown that this reflection symmetry (as well as time-reversal symmetry) is preserved for any quantum walk where $\phi_2-2 \phi_1=0~ (\text{mod}~ 2 \pi)$ \footnote{This can be seen by performing the gauge transformation $\ket{j} \rightarrow e^{- \text{i} j \phi_1}\ket{j}$ in Eq.~\eqref{eq:QW_NNN}}. A possible way to break this reflection symmetry and bias the walk is to introduce imaginary NNN couplings. This is in contrast with the 1D discrete-time quantum walk where even when it is possible to hop only between NN, an asymmetric propagation of an initially localized state can be achieved either by tuning the initial coin state or by appropriately choosing the coin-flip operator \cite{kendon_controllingDTQW}.     

The protocol we summarize in Table~\ref{table:realandeffHams} simulates the time evolution of a quantum walk with Hamiltonian $H_{1D}(|K_1|, \im |K_2|)$. In order to achieve an evolution time $T_\text{evol}$, we can set $J^{(0)}= 3 K_1 T/(4 \pi)$, $J^{(1)}= \sqrt{|K_2|}$. This choice implies that the effective Hamiltonian becomes  
\begin{equation}
    H_{\eff}\approx \frac{3 T }{4 \pi} H_{1D}(|K_1|, \im |K_2|),
\end{equation}
which is a rescaled version of $H_{1D}(|K_1|, \im |K_2|)$. Hence, the time needed to simulate the time evolution of $H_{1D}(|K_1|, \im |K_2|)$ for time $T_{\text{evol}}$ using $H_{\eff}$ is given by 
\begin{equation}
    T_\text{sim}= \frac{4 \pi T_\text{evol}}{3 T}.
\end{equation}
Note that the period $T$ has to be sufficiently small to ensure a good simulation accuracy. This implies that, in general, the simulation time is larger than $T_\text{evol}$.  

We have verified numerically that the choice $K_1=1$ and $K_2=0.2 \im$ leads to a particularly asymmetric propagation of the quantum walk. In Fig.~\ref{fig:QW_imNNN}, we show the probability distribution of such a walk after a fixed time and compare it to that obtained via our simulation protocol, observing a good agreement even for values of the period as large as $T=0.5$. The dynamic of this CTQW with imaginary NNN couplings is characterized by a sharp peak propagating to the right, with smaller waves propagating to the left at a faster speed. This is in clear contrast with the dynamics of the quantum walk with real NNN couplings of the same strength ($K_1=1$,  $K_2=0.2$), also shown in Fig.~\ref{fig:QW_imNNN}. The latter preserves time-reversal symmetry, and the probability distribution exhibits a reflection symmetry around the initial position of the walk. It is thus clear that changing the phase of $K_2$ can have a striking effect on the propagation.    

We also remark that the dispersion relation of the 1D quantum walk with Hamiltonian $H_{1D}(|K_1|, \im |K_2|)$, assuming periodic boundary conditions, is given by
\begin{align}
    E_k = 2 |K_1| \cos{(2\pi k/N)}-2 |K_2| \sin{(4\pi k/N)},
\end{align}
where $k\in \{0,1,..,N-1\}$ labels the momentum eigenstates. Our protocol can thus be used to simulate the propagation of particles in 1D systems with various engineered dispersion relations.
\begin{table*}
\def\arraystretch{2.5}
\centering
\begin{tabular}{M{2.5cm} p{11cm} M{3.5cm} }
\hline \hline
\textbf{Hamiltonian} & & \textbf{Connectivity} \\
\hline \hline
real & $H(t)=\sum_{j=1}^{N-1}\left[ J^{(0)}+J^{(1)} \; \left(\cos(\Omega t-j\pi)-2\cos(2\Omega t-j\pi )\right)\right]\ket{j}\bra{j+1}+\text{h.c.}$ & 1D with NN couplings \\ 
effective & $H_{\eff}\approx \sum_{j=1}^{N-1} J^{(0)}\ket{j}\bra{j+1}+ \sum_{j=1}^{N-2}  \frac{3 \im }{2 \Omega} (J^{(1)})^2 \ket{j}\bra{j+2} +\text{h.c.}$ & 1D with NN + NNN couplings \\
\hline
\hline
real  & $H(t)=\sum_{j\in P} J_{0j}\cos(\Omega t)\ket{0}\bra{j}+\sum_{j\in \overline{P}}J_{0j}\cos(\Omega t+\pi/2)\ket{0}\bra{j}+\text{h.c.}$        & star graph   \\
effective &  $H_{\eff}\approx \frac{1}{2\Omega} (J^{(1)})^2\sum_{j\in P} \sum_{k\in \overline{P}} \ket{j}\bra{k}+\text{h.c.}$ & complete bipartite graph \\
\hline
\hline
    \end{tabular}
\caption{On top, we summarize the protocol to simulate uniform imaginary NNN couplings in a quantum walk in the linear chain from a one-dimensional quantum walk with time-dependent NN couplings. On the bottom, we summarize the protocol to simulate a quantum walk on a complete bipartite graph from a time-dependent quantum walk on the star graph. The frequency $\Omega$ has to be large enough compared to the other energy scales for the effective description to be accurate.  \label{table}}
\label{table:realandeffHams}
\end{table*}

\subsection{Simulating a quantum walk on a complete bipartite graph}\label{sec:sim_CBG}

\begin{figure}
\centering
\includegraphics{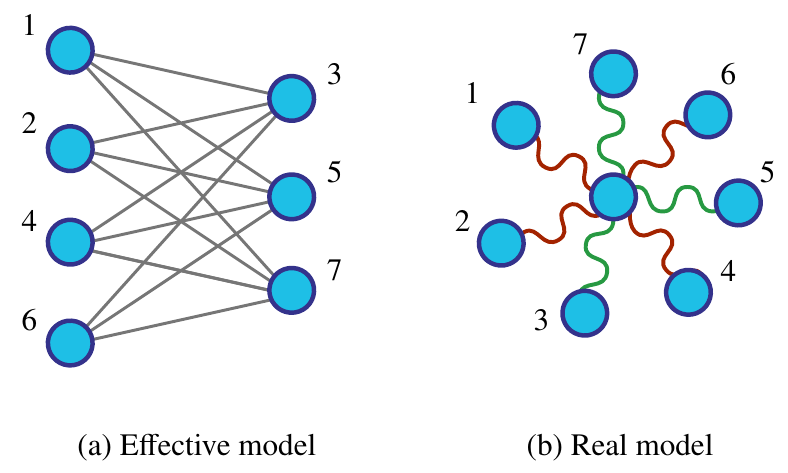}
\caption{(Color online) a) Complete bipartite graph of 7 nodes, with partitions $P=\{1,2,4,6\}$ and $\bar{P}=\{3,5,7\}$. b) Schematic representation of the protocol to simulate a quantum walk on a complete bipartite graph via time-periodic modulation of the couplings of a Hamiltonian with a star connectivity (all nodes coupled to a single central node). The partition $P$ and $\overline{P}$ can be chosen via two different modulations of the oscillating phase, represented by the red and green colors.} 
\label{fig:CBG}
\end{figure}
In the previous example, time-periodic coupling modulation was used to effectively introduce NNN couplings in 1D quantum walks, increasing the number of edges by a factor of 2. Here, we demonstrate that if we start with a graph with different connectivity, for instance, a star graph, it is possible to engineer quantum walks on considerably more complex graphs, where the number of edges can be effectively increased by a factor of $N$. More precisely, we will demonstrate a protocol to simulate a quantum walk on a complete bipartite graph of $N$ nodes, which can have up to $\mathcal{O}(N^2)$ edges, starting from a time-dependent quantum walk on a graph with star connectivity of $N+1$ nodes. 

Let the central node of a star graph in Fig.~\ref{fig:CBG}~b) be denoted as $\ket{0}$, and consider the Hamiltonian 
\begin{align}\label{eq:Hstar}
    H(t)&=\sum_{j=1}^N J_{0 j}(t) \; (\ket{0}\bra{j}+\ket{j}\bra{0}),
\end{align}
where $J_{0 j}(t+T)=J_{0 j}(t)$. In this case, we assume that only the first frequency $\Omega=2\pi/T$ is present in the Fourier expansion (see Eq.~\eqref{eq:couplingsFourier}). This implies that the zeroth order term of the Magnus expansion vanishes. For simplicity, we choose a uniform value for $J^{(1)}_{0 j}=J^{(1)}$, but our results are valid in the more general scenario of variable coupling strength. Using Eq.~\eqref{eq:NNNcouplings}, we obtain that 
\begin{align}\label{eq:Heffstar}
    H_{\eff}^{(1)}= \frac{\im}{2 \Omega} (J^{(1)})^2\sum_{j=1}^N\sum_{k=1}^N \sin \pc{\phi^{(1)}_{0,j}-\phi^{(1)}_{0,k}}\ket{j}\bra{k}. 
\end{align}
The resulting Hamiltonian only contains couplings between the nodes $\{1,2,\dots,N\}$, i.e., the central node $0$ gets effectively decoupled. Furthermore, the choice of the configuration of the phases $\phi^{(1)}_{0,j}$ determines which nodes are effectively connected. 

The connectivity of a complete bipartite graph with uniform couplings can be obtained as follows. Let $P$ be a subset of the set of all nodes $\{1, 2, \dots N\}$, and $\overline{P}$ its complement. A complete bipartite graph is one where each node from subset $P$ is connected to all other nodes from $\overline{P}$ and vice-versa (see Fig.~\ref{fig:CBG}~a)). 
Choosing the phases $\phi^{(1)}_{0,j}=\pi/2$ if $j\in P$, and $\phi^{(1)}_{0,j}=0$ if $j\in \overline{P}$, we obtain from Eq.~\eqref{eq:Heffstar} the effective Hamiltonian
\begin{align}\label{eq:CBGcomplex}
  H_{\eff}^{(1)}= \frac{\im}{2\Omega} (J^{(1)})^2\sum_{j\in P} \sum_{k\in \overline{P}} \pr{\ket{j}\bra{k}-\ket{k}\bra{j}},
\end{align}
which represents the Hamiltonian of a quantum walk on a complete bipartite graph, with an imaginary effective coupling strength $\frac{\im}{2 \Omega} (J^{(1)})^2$. However, in this case, the fact that the coupling is imaginary has no observable physical consequences, since this Hamiltonian can be transformed into a real Hamiltonian via a gauge transformation. More precisely, if we define $\ket{k'}=\im \ket{k}$, for $k\in \overline{P}$, we can rewrite Eq.~\eqref{eq:CBGcomplex} as 
\begin{align}
 H_{\eff}^{(1)}= \frac{(J^{(1)})^2}{2\Omega} \sum_{j\in P} \sum_{k'\in \overline{P}} \pr{\ket{j}\bra{k'}+\ket{k'}\bra{j}}.   
\end{align}
Our scheme to simulate the quantum walk on the complete bipartite graph is summarized in Table~\ref{table:realandeffHams}. This example highlights the fact that the original connectivity of the graph can be significantly increased by using the freedom of periodic modulation of the couplings. Furthermore, although we have focused on a simple choice of phases that leads to the simulation of the complete bipartite graph's connectivity, we believe that more complex choices of phases in Eq.~\eqref{eq:Heffstar} can be exploited towards the simulation of quantum walks on many other different families of graphs. 

\subsection{Experimental feasibility}\label{sec:exp_NNN}
\begin{figure}[t!]
\centering
\includegraphics{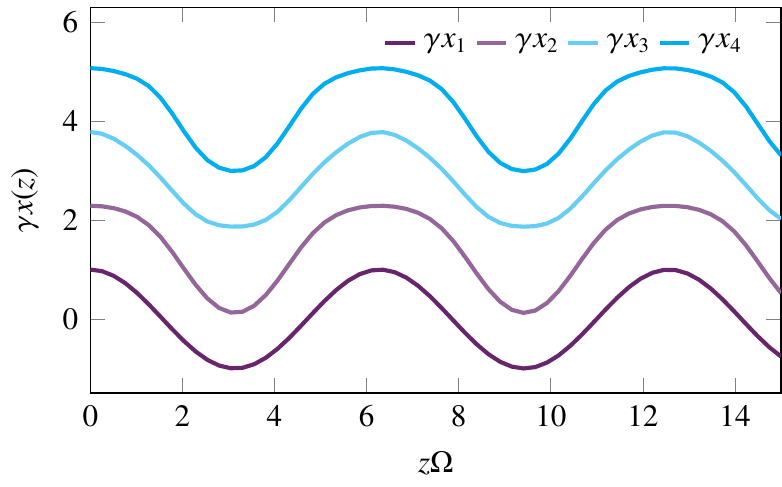}
\caption{(Color online) Proposed set of functions $x_j (z)$ required to perform the protocol for simulating imaginary NNN coupling in a system of coupled waveguides, where couplings decrease exponentially with the distance. We have set $\gamma x_1 (z) = \cos \pc{\Omega z}$, and the following $x_j (z)$ are found by solving Eq.~\eqref{eq:wg} with $\kappa=4$, $J^{(0)}=1$ and $J^{(1)}=0.1$.
\label{Fig:IPCexp} }
\end{figure}

 For the reasons already mentioned in Sec.~\ref{sec:experimentscomplexphases}, an ideal experimental platform to implement our simulation protocol for a quantum walk with imaginary NNN couplings are integrated photonic circuits \cite{szameit_review}.  The high degree of control of the spatial configuration of laser-written waveguides can be exploited to create the desired form for the periodic couplings of the time-dependent Hamiltonian discussed in Sec.~\ref{sec:1d_imNNN} (see also Table~\ref{table}). Specifically, let us assume that the waveguides are written in $xz$-plane, where the $z$-direction corresponds to the propagation direction of the photon, which plays the role of time, whereas the relative distance between two waveguides is given by $|x_j(z)-x_{j+1}(z)|$.  As the evanescent coupling between two adjacent waveguides decreases exponentially with their relative distance, we consider an ansatz for the coupling of the form  $J_{j, j+1} = \kappa \exp \pr{- \gamma (|x_j(z) - x_{j+1}(z)|) }$. In Fig.~\ref{Fig:IPCexp}, we represent a set of functions $x_j(z)$ such that the $z$-dependence of the couplings has the form required by the protocol for simulating imaginary NNN couplings of Sec.~\ref{sec:1d_imNNN}, i.e., they obey the relation
\begin{align}
\kappa e^{- \gamma |x_j(z) - x_{j+1}(z)|} &= J^{(0)} +J^{(1)}\left[\cos(\Omega z-j\pi)-2\cos(2\Omega z-j\pi )\right].
\label{eq:wg}
\end{align}
Here, we have chosen the first waveguide's shape to be $x_1(z)=\cos(\Omega z)$. By modulating the waveguides in this way and provided the modulation frequency $\Omega$ is fast enough, it is, in principle, possible to observe the effects of asymmetric propagation of the quantum walk with imaginary NNN couplings in this experimental platform. 

We note that one-dimensional CTQWs with real NNN interactions have been previously implemented in arrays of photonic crystals~Ref.~\cite{arxiv:1611.02520}. In contrast, our results from Sec.~\ref{sec:1d_imNNN} allow for the implementation of quantum walks of the form of Eq.~\eqref{eq:QW_NNN} with imaginary NNN couplings, which break time-reversal symmetry. In addition, we remark that such a Hamiltonian can describe the dynamics of charged particles on a triangular ladder with a staggered magnetic field and is thus of interest in the context of condensed matter physics~\cite{Cabedo_2020}. A proposal to simulate such a system, motivated by its many-body properties in the presence of interactions, was put forward in Ref.~\cite{Cabedo_2020} and exploits light-induced spin-orbit coupling on optical lattices of ultracold atoms.  On the other hand, our work focuses on single particle dynamics, which is easier to implement and control via photonic quantum walks. Furthermore, an attractive feature of our simulation method is that it does not require the coupling of spatial and internal degrees of freedom as in Ref.~\cite{Cabedo_2020} and only assumes time dependent control over the coupling strength between different nodes. 

Regarding the proposal from Sec.~\ref{sec:sim_CBG}, the main challenge is to implement a Hamiltonian with the connectivity of a star graph. This is hard to obtain in systems where the couplings depend on the distance between different nodes of the quantum walk, especially for quantum walks with a large number of nodes. However, we remark that bosonic hopping Hamiltonians with the connectivity of a star graph can be realized with superconducting qubits by coupling different oscillators to the same so-called ``bus" mode~\cite{Floquet_annealers}. Hence, our scheme to simulate the quanutm walks on complete bipartite graphs with tunable connectivity could, in principle, be realized if it is possible to modulate the coupling coefficients at high enough frequency. 

\section{Conclusion} \label{sec:conclusions}
Floquet theory allows us to describe the dynamics of periodically driven systems in terms of an effective Hamiltonian. We use this idea in the context of continuous-time quantum walks (CTQWs): by fast periodic control over certain parameters of a quantum walk on a given graph, an effective description emerges, which can be seen as a quantum walk on a new graph with renormalized coupling parameters. This opens up the possibility to engineer new types of quantum walks that could not be implemented directly due to experimental restrictions. 

We have investigated two main applications of Floquet engineering to CTQWs. Firstly, we have discussed how fast periodic modulation of on-site energy terms can be exploited to implement effective complex coupling coefficients. This can be used to control the propagation direction of the walker and improve transport efficiency in specific graphs \cite{ScienRep3.2361(2013)}. 
Secondly, we have shown how effective NNN couplings can be introduced via time-dependent control over the coupling coefficients, leading to simulation protocols for quantum walks on highly connected structures. The presence or absence of these NNN couplings can be controlled through the shape of the modulation functions, and their strength can be tuned to be as large as that of the NN couplings.

The effective NNN couplings obtained by our approach are purely imaginary and can lead to time-reversal symmetry breaking. This effect is evident in the quantum walk dynamics in the one-dimensional chain with imaginary NNN couplings, which is one example we explicitly show how to simulate.  Our work may inspire experimental implementations of robust protocols achieving high fidelity quantum state transfer such as the one introduced in Ref.~\cite{SSH_imNNN}, which requires purely imaginary tunable  NNN couplings in a Su-Schrieffer-Heeger chain. Furthermore, it would be interesting to extend our approach to two and three-dimensional lattices. This could lead to new ways of engineering dispersion relations or exploiting time-reversal symmetry breaking in higher-dimensional quantum walks, possibly leading to improvements in the performance of state transfer protocols or search algorithms on regular lattices \cite{childs2004spatial, tanner_search}. We also leave open the question of whether \emph{real} NNN couplings can be effectively implemented in quantum walks via Floquet engineering. In principle, this could be used to simulate quadratic perturbations to the usual quantum walk Hamiltonian~\cite{parisNNN} or quantum walks on other structures such as complex networks~\cite{biamonte2019complex}. 

We remark that after the completion of this work, the long time dynamics of the one-dimensional quantum walk with complex NNN couplings were analyzed in Ref.~\cite{bhandari_imaginaryNNN}, extending previous work for real NNN couplings~\cite{bhandariNNN}. This work shows that the propagation of the walk exhibits different velocities between right and left-moving wavefronts and that the skewness of the distribution is maximal for purely imaginary NNN couplings.

Overall, our work opens up new possibilities for exploiting time-dependent control of quantum walk Hamiltonians towards directing and enhancing quantum transport in networks as well as investigating quantum dynamics in highly connected structures.

\section*{Acknowledgments}
We thank Shantanav Chakraborty, Hugo Ter\c{c}as, Michael Steel, Diego Guzman Silva and Fulvio Flamini for discussions. SR is supported by the EPSRC grant EP/R002061/1. LN acknowledges support from F.R.S-FNRS. 
%
\appendix
\section{Alternative form for the modulation functions} \label{app:alternative_modulation}
Following the work of Refs.~\cite{struck2012, NatPhys9.738(2013)}, we consider the set of modulation functions
\begin{align}
\beta^{(\sin)}_2(t)&=\begin{cases}A \sin\left(\dfrac{3\pi t}{T}\right),~0\leq t \leq \frac{2T}{3},\label{eq:beta2sin}\\~\\
0,~ \frac{2T}{3}< t\leq T,\end{cases} \\ 
\beta^{(\sin)}_3(t)&=\begin{cases}0,~0\leq t \leq \frac{T}{3},\\~\\ 
A \sin\left(\dfrac{3\pi t}{T}\right),~ \frac{T}{3}< t\leq T.\end{cases}\label{eq:beta3sin}
\end{align}
for the site energies of a quantum walk on the triangular loop discussed in Sec.~\ref{subsec:triangle}. These functions are lead to the effective couplings
\begin{align}
J^{\text{eff}}&=J_{12}^{\text{eff}}=J_{23}^{\text{eff}}=J_{31}^{\text{eff}} \nonumber\\
&=J'\left[\frac{2}{3}J_0\left(\frac{AT}{3\pi}\right)e^{\im \frac{AT}{9\pi}}+\frac{1}{3}e^{\im \frac{2AT}{9\pi}}\right]\label{eq:eff_coupling},
\end{align}    
where we have used Eq.~\eqref{eq:Jeff} and the integral relation
\begin{align*}
\int_0^{\pi} d \theta e^{\im x \cos(\theta)}&=\pi J_0(x),\;
\end{align*}
with $J_0(x)$ representing the zero-order Bessel function. 

\begin{figure}[h]
    \centering
    \includegraphics{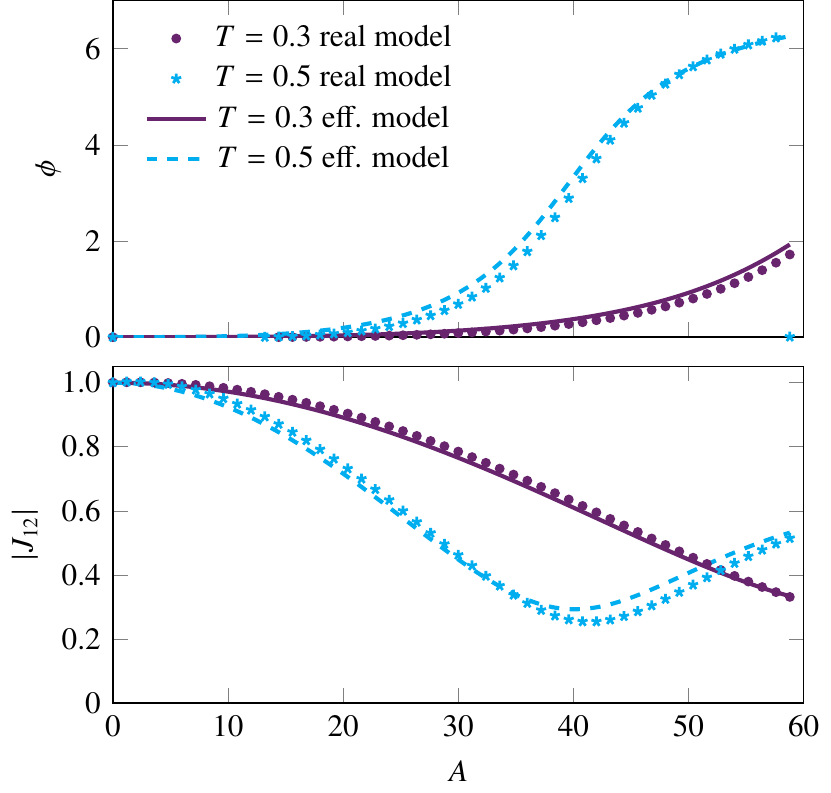}
    \caption{(Color online) Comparison between real and effective models for the accumulated phase around the triangular loop $\phi$ (top) and absolute value of coupling $|J^{\eff}_{12}|$ (bottom) for two values of the period $T=0.3$ (in purple) and 0.5 (in blue) as a function of the modulation amplitude $A$, for the sine-like modulation functions from Eq.~\eqref{eq:beta2sin} and \eqref{eq:beta3sin}. The theoretical values were extracted from Eq.~\eqref{eq:eff_coupling} and the value of $J'$ was set to 1.}
    \label{fig:comparison_realvseff_sine}
\end{figure}

In Fig.~\ref{fig:comparison_realvseff_sine}, we compare the parameters $|J^\text{eff}|$ and $\phi$, extracted from a numerical calculation of $H_{\eff}=\im \log(U(T))/T$ and from the analytical calculation from Eq.~\eqref{eq:eff_coupling}, showing a good agreement between the two for the range of parameters shown.    
It is interesting to observe that, the step-like functions lead to higher effective phases $\phi$ for the same values of $A$ and $T$, when compared to the sine-like modulation functions. For example, if we fix the modulation period at $T=0.5$, a value of the effective phase of $\pi/2$ can be achieved for an amplitude $A\approx 26$ (in units where $J'=1$) using the step-like functions (see Fig.~\ref{fig:comparison_realvseff_step}), whereas for the sine-like modulation we would need an amplitude $A\approx 35$ to obtain the same effective phase. This type of considerations is of paramount importance for experimental realizations, where there are natural or technological limits to the possible modulation amplitudes that can be reached (see Subsec.~\ref{sec:experimentscomplexphases}).

\bibliography{Bibliography}

\end{document}